\documentclass[12pt]{emulateapj}
\newcommand{\beq}{\begin{equation}}
\newcommand{\eeq}{\end{equation}}

\newcommand{\hi}{H{\sc i}~}

\newcommand{\hia}{H{\sc i}}

\usepackage[colorlinks,urlcolor=blue,citecolor=blue,linkcolor=blue]{hyperref}
\usepackage{gensymb}
\usepackage{amsmath}
\usepackage{color}


\newcommand{\citei}[1]{\citeauthor{#1} \citeyear{#1}}

\newcommand{\kms}{km ${\rm s^{-1}}$}

\begin{document}

\title{Magnetically Aligned HI Fibers and the Rolling Hough Transform}
\author{Clark, S.E.\altaffilmark{1,}}
\author{Peek, J.E.G.\altaffilmark{1,}\altaffilmark{2}}
\author{Putman, M.E.\altaffilmark{1}}
\altaffiltext{1}{Department of Astronomy, Columbia University, New York, NY} 
\altaffiltext{2}{Hubble Fellow}

\begin{abstract}
We present observations of a new group of structures in the diffuse Galactic ISM: slender, linear \hi features we dub ``fibers'' that extend for many degrees at high Galactic latitude. To characterize and measure the extent and strength of these fibers, we present the Rolling Hough Transform (RHT), a new machine vision method for parameterizing the coherent linearity of structures in the image plane. With this powerful new tool we show the fibers are oriented along the interstellar magnetic field as probed by starlight polarization. We find that these low column density ($N_{HI} \simeq  5 \times 10^{18}$ cm$^{-2}$) fiber features are most likely a component of the local cavity wall, about 100 pc away. The \hi data we use to demonstrate this alignment at high latitude are from the Galactic Arecibo L-Band Feed Array \hi (GALFA-\hia) Survey and the Parkes Galactic All Sky Survey (GASS). We find better alignment in the higher resolution GALFA-\hi data, where the fibers are more visually evident. This trend continues in our investigation of magnetically aligned linear features in the Riegel-Crutcher \hi cold cloud, detected in the Southern Galactic Plane Survey (SGPS). We propose an application of the RHT for estimating the field strength in such a cloud, based on the Chandrasekhar-Fermi method. We conclude that data-driven, quantitative studies of ISM morphology can be very powerful predictors of underlying physical quantities.
\end{abstract}


\section{Introduction}

Magnetic fields, radiation, turbulence, and cosmic rays are major players that mold the diffuse interstellar medium (ISM). The prevalence of starlight photons and cosmic rays partially ionizes the largely neutral medium, and causes magnetic fields and gas to move together (i.e., flux freezing). We therefore expect the geometry and strength of the interstellar magnetic field to affect the shape of the ISM. Studies of the magnetic field in diffuse \hi (n ${\sim}$ 0.1 - 100 cm${^{-3}}$) suggest that the field strength is relatively independent of volume density, in contrast to magnetic fields in molecular clouds (e.g., \citei{Heiles:2005tb}). The role of magnetic fields in molecular cloud and star formation is an area of active research (see \citei{Crutcher:2012hw} for a recent review). A better understanding of the magnetic structure of the diffuse ISM, the medium from which denser structures form, may elucidate the processes at work on all scales. 

Sensitive, high spatial dynamic range \hi observations allow us to observe the structure of the diffuse ISM in unprecedented detail. These observations have resolved the previously ``blobby'' ISM into a complex network of filaments, clumps, and shells. Even a cursory inspection of these data indicates that the ISM is not a simple, self-similar, turbulent medium easily described by a few parameters, but rather an enormously complex structure affected by many discrete processes on a wide range of scales.
Traditionally, such features within the ISM have been identified by eye \citep[e.g.,][]{Mc-CG06b, 2010ApJ...722..395B}, though there have been some attempts to automate the process for relatively simple structures \citep[e.g.,][]{Saul:2012gt}.  Many numerical investigations of ISM data have revolved around functions that either strip out Fourier phase information and rely heavily on power spectra, or examine the hierarchical clustering of gas \citep[e.g.,][]{Burkhart:2011ui}. Some work has been done to build metrics that quantify morphology \citep{Adams:1992kh,KJN04,Robitaille:2010ii}, though these metrics are designed to be general, rather than to capture information about specific observed features.  There are very few methods that have quantified shape information in the ISM and use it as a predictor of an underpinning physical property.

\begin{figure*}
\centering
\includegraphics[scale=0.9]{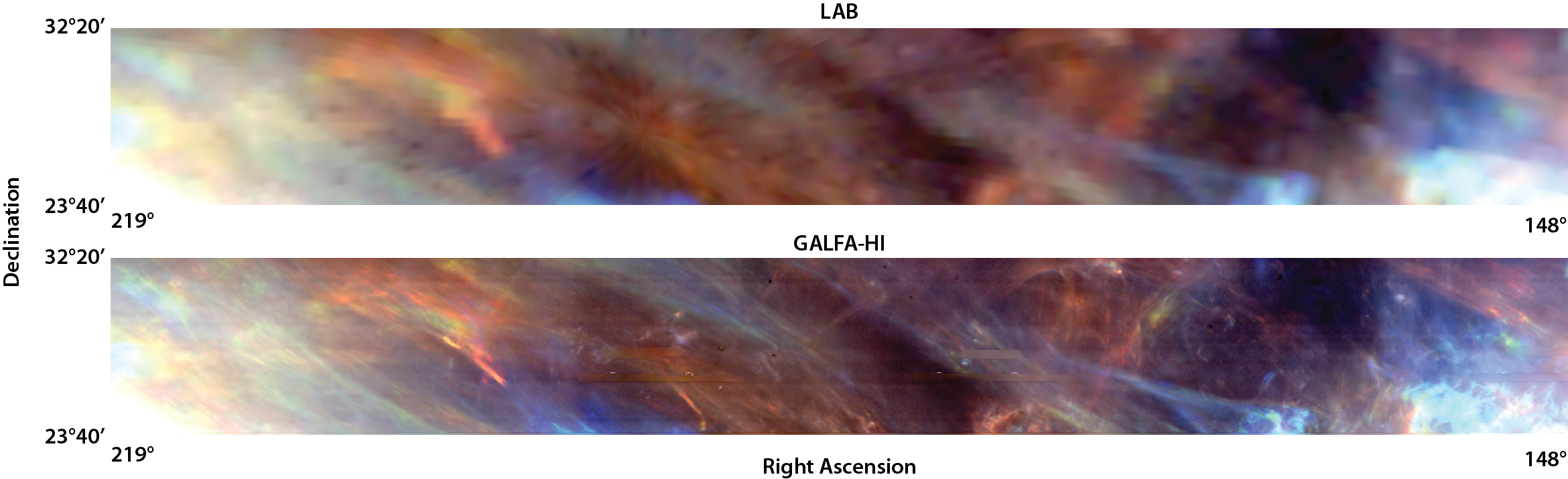}
\caption{\hi data at high Galactic latitude. Top panel is taken from the 36$^\prime$ resolution Leiden-Argentina-Bonn survey \citep[LAB]{Kalberla05}, bottom panel from a section of the 4$^\prime$ resolution GALFA-\hi DR1 data analyzed in this work. Red, blue, and green channels represent -7 to -4 \kms, -3 to -1 \kms, and 0 to 3 \kms, respectively. Brightnesses are shown in a logarithmic stretch in brightness temperature from 0.5 K (dark) to 5 K (light), or an \hi column density range of $3 \times 10^{18}$ cm$^{-2}$ to $3 \times 10^{19}$ cm$^{-2}$. The slender fiber features can be seen in the bottom panel but are washed out by low resolution of the LAB survey in the top panel.}
\label{fiberfig}
\end{figure*}

The Galactic Arecibo L-Band Feed Array \hi (GALFA-\hia) Survey  is mapping 13,000 square degrees of sky at 4${\arcmin}$ resolution. At this high spatial resolution, we observe that the diffuse, high-latitude \hi is organized into high aspect ratio structures we call fibers (Figure \ref{fiberfig}). We often find them in groups largely parallel to each other. We use the term ``fibers" to evoke the slender, parallel nature of these \hi features. They are visually similar to slender molecular fibers identified in star forming regions (e.g., \citei{2013arXiv1312.6232A}; \citei{Hacar:2013uh}). While the term ``filaments" is used in the literature to describe a wide range of linear structure, we reserve it in this work to refer to networks of gravitationally bound structures found by other authors. 

Why does such striking linear structure pervade the high-latitude ISM? The elongation we see in these fibers suggests that magnetic fields may play a crucial role in determining the structure of the diffuse ISM. In this work we explore the correlation between the orientation of the magnetic field, as traced by starlight polarization, and the orientation of these gaseous fibers. Starlight polarization traces the orientation of the plane-of-sky magnetic field because the starlight is polarized by magnetically aligned interstellar grains (\citei{Goldsmith:2008hl}). To examine this correlation quantitatively, we require a method for detecting and parameterizing linear structure. In \S \ref{rht} we develop a machine vision algorithm, the Rolling Hough Transform (RHT), designed for this purpose. This powerful new technique allows us to quantify the alignment of \hi fibers with the magnetic field using diagnostics we develop in \S\ref{starpolmethods}. In \S\ref{data} we detail the data used in this study. We investigate the gas-magnetic field alignment in diffuse \hi in \S\ref{diffuse}, and apply the same analysis to the Riegel-Crutcher \hi self-absorption feature in \S\ref{rccloud}. The success of the RHT at mapping the detailed structure of the magnetic field in the Riegel-Crutcher cloud suggests a technique for resolved field strength estimation, which we propose in \S\ref{cfmethod}. We discuss the implications of the work in \S \ref{discussion} and conclude with a summary and prospects for future work in \S \ref{conclusion}.

\section{The Rolling Hough Transform (RHT)} \label{rht}

The detection of astronomical linear structure is approached in various ways depending on the context. In cosmic web data, filaments are described as structures linking local density maxima (e.g. the DisPerSE method of \citei{Sousbie:2011ft} and the SHMAFF method of \citei{Bond:2010ct}). DisPerSE has also been used in the context of filaments in the molecular ISM, as in the \emph{Herschel} filaments analyzed in \cite{Arzoumanian:2011wu}. A rich methodology for linear and curvilinear feature detection has been developed for analysis of solar data (see \citei{Aschwanden:2009gk} for a broad review of solar image processing and feature detection). The curvelet transform as described by \citet{Starck:2003wx} has been used across a number of sub-disciplines to highlight and enhance linear features in astronomical images. \citeauthor{Hennebelle:2013ws} (\citeyear{Hennebelle:2013ws}) uses the inertia matrix to isolate filaments in simulation data.

We wish to quantify the linearity and spatial coherence of \hi structures. Because these structures are not objects with distinct boundaries (see Figure \ref{fiberfig}), we are tackling a problem that is fundamentally different from solar feature detection and filament identification. Additionally, the filament detection algorithms used in solar observations (and \citei{Starck:2003wx}) report images as their results, which do not directly produce a quantitative measure of linearity in the image. As these diffuse \hi fibers were not formed by gravitational forces, there is no reason to require that they must be, or bridge, local overdensities. Indeed, we find the fibers often to be in groups of parallel structures, very unlike the cosmic web. Thus, methods developed for gravitationally-dominated systems are not optimal for our purposes.

The RHT is, as its name suggests, a modification of the Hough transform.  The Hough transform was first introduced in a patent for the detection of complex patterns in bubble chamber photographs (\citei{Hough:1962tb}).  It was soon recognized as a powerful line detection technique, and has found wide applications in image processing and machine vision (for an excellent review, see Illingworth and Kitler \citeyear{Illingworth:1988uo}). 
The adaption of the Hough transform described here is a rolling version that is particularly well suited to the detection and quantization of specific linear features in astronomical data.
The RHT does not merely identify fibers; it encodes the probability that any given image pixel is part of a coherent linear structure. This allows the user to quantify the linearity of regions of sky without specifying fibers as discrete entities.

\subsection{RHT Procedure}\label{rhtproc}

\begin{figure*}
\centering
\includegraphics[scale=0.5]{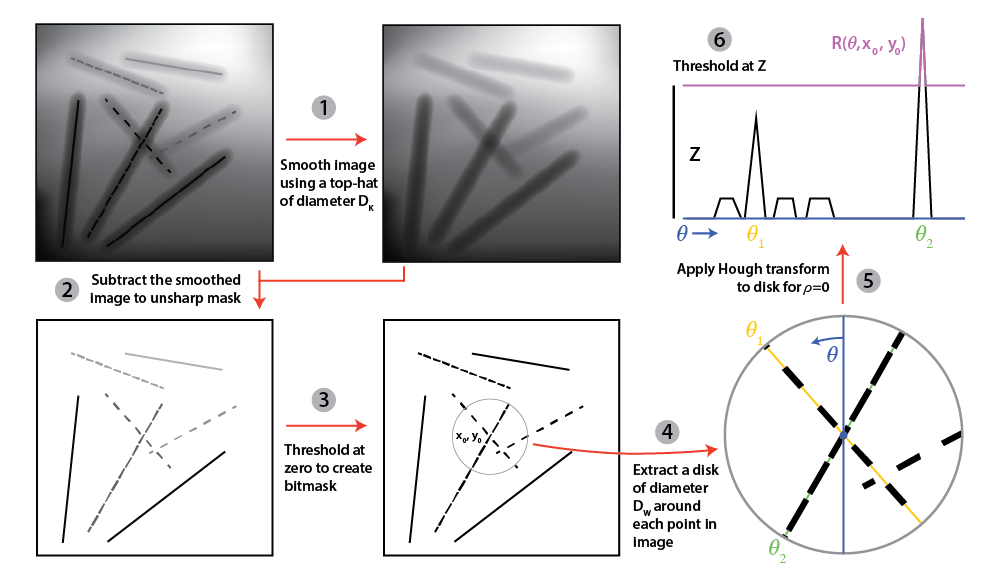}
\caption{A diagram of the RHT procedure (\S\ref{rhtproc}). Steps 1-3 are preprocessing of the image. Step 4 shows the selection of a disk of diameter $D_W$. This window rolls across the data, centered on each pixel in turn. Step 5 shows the Hough transform applied to cartoon data, and step 6 illustrates that only data above a defined threshold is recorded. Note that this cartoon data contains three linear features, two of which (green and yellow) are centered on the selected window center $(x_0, y_0)$, and contribute the most intensity to the Hough transform. The dashed lines are representative of different levels of coherence in the data. Here, only the green line (with $\theta = \theta_2$ orientation) has RHT intensity $R(\theta, x_0, y_0)$ over the threshold Z.}
\label{diagram}
\end{figure*}

The RHT operates on two-dimensional data and is designed to be sensitive to linear structure irrespective of the overall brightness of the region. The first step is to unsharp mask the image. The image is convolved with a two-dimensional top-hat smoothing kernel of a user-defined diameter, $D_K$ (Figure \ref{diagram}, step 1). The smoothed data is then subtracted from the original data (Figure \ref{diagram}, step 2), and the resulting map is thresholded at 0 to obtain a bitmask (Figure \ref{diagram}, step 3). The subtraction of the smoothed component can be considered a suppression of large-scale structure, or a high-pass Fourier filter.

Our implementation of the Hough transform follows that of Duda and Hart (\citeyear{Duda:1972uj}), where a straight line is parameterized in terms of the angle $\theta$ of its normal, and its minimum Euclidean distance from the origin, $\rho$:
\beq
\rho = x\cos\theta + y\sin\theta. 
\eeq
\noindent This parameterization avoids the computationally problematic singularities that can arise in a point-slope description of a line.

Every possible line in the image space is uniquely specified by a point in the ${\rho}$-${\theta}$ space. The standard Hough transform maps each $(x, y)$ pixel in the image space to all (${\rho}$, ${\theta}$) line parameters possible for that pixel in the $\rho$-$\theta$ space. The Hough transform is thus a one-to-many mapping between image space and parameter space. The Hough transform stores in a (${\rho}$, ${\theta}$) ``accumulator array'' the number of ``on'' pixels in image space that contribute to each pixel in the $\rho$-$\theta$ space. All values in the (${\rho}$, ${\theta}$) accumulator array over a set threshold are then identified as a line in the image space.

The RHT performs a similar mapping from image space to parameter space, with several key differences. The RHT mapping is performed on a circular domain, diameter $D_W$, centered on each image-space pixel $\left(x_0, y_0\right)$ in turn (Figure \ref{diagram}, step 4). Then a Hough transform is performed on this area, limited to  ${\rho}$ = 0 (Figure \ref{diagram}, step 5). Thus the $\rho$-$\theta$ space is reduced to a one-dimensional space on $\theta$ for each pixel. All intensity over a set intensity threshold $Z$ is stored as $R(\theta, x_0, y_0)$: RHT intensity as a function of $\theta$ for that pixel (Figure \ref{diagram}, step 6). $Z$ is a percentage. In every direction ${\theta}$, $Z \times D_W$ pixels must contain signal in order for the transform to record the data in that direction.
We use the canonical binning for the number of theta bins:
\beq
n_{\theta} = \left\lceil\pi \frac{\sqrt{2}}{2} \left(D_W - 1\right) \right\rceil
\eeq
The mapping of each pixel in the circular region to the reduced domain (${\rho}$ = 0, ${\theta}$) is defined by the Hough transform. 
As the Hough transform is distributive over image coadditon, we tabulate this mapping in advance for each pixel within the circular region to optimize the RHT. By iterating (``rolling'') over the entire image space we produce the RHT output, $R\left(\theta, x, y\right)$.
A visualization of the linear structures identified by the RHT, the backprojection $R\left(x, y\right)$, is obtained by integrating ${R(\theta, x, y)}$ over ${\theta}$:
\beq\label{Bxy}
R(x, y) = \int R(\theta, x, y)\,d\theta.
\eeq

The bottom panels of Figures \ref{galfareg} and \ref{gassreg} show RHT backprojections.

\subsection{Parameter Space}\label{pspace}

One advantage of the RHT is that the input parameters of the transform can be chosen to highlight specific linear features of interest. One defines, for a given run of the RHT, a smoothing kernel diameter ($D_K$), window diameter ($D_W$), and intensity threshold ($Z$), as described above. The rolling nature of the RHT ensures that linear structure at least as long as $D_W$ will be identified. Thus $D_W$, along with the $Z$, sets a lower limit for the spatial length of the linear features. Thresholding below ${100\%}$ ($Z < 1$) reflects the fact that structures can be physically coherent even if they are not visibly connected (see Figure \ref{diagram}). With Galactic \hi data we have radial velocity as well as spatial information, so we choose a specific velocity ($v$) and velocity range ($\delta v$) to generate an image on which to run the RHT.

\section{RHT-Starlight Polarization Methods}\label{starpolmethods}
We describe two metrics for quantifying the degree of alignment between RHT output, hereafter ${R(\theta, x, y)}$, and starlight polarization angle, hereafter ${\theta^\star}$. $R(\theta, x, y)$ is intensity as a function of angle on a domain ${\theta \in [0, \pi)}$, as a ${0 \degree}$ orientation is equivalent to a ${180 \degree}$ orientation. Similarly, ${0 \degree}$ and ${180 \degree}$ are equivalent starlight polarization angles.

In what follows, we sample ${R(\theta, x, y)}$ in a circular region around each star in the field:
\beq
R_\star(\theta)  = \iint \limits_{disk}  R(\theta, x, y)  \,dxdy.
\eeq
 
We visualize this on a half-polar plot, such that perfect alignment between ${R_\star(\theta)}$ and ${\theta^\star}$ lies at 0, and orthogonal alignment lies at ${\theta = \pi/2}$ or ${-\pi/2}$. This amounts to shifting $R_\star\left(\theta\right)$ to $R_\star\left(\phi\right)$, where:
\beq
\phi \equiv \theta - \theta^*,
\eeq
 and this subtraction occurs on the domain $\theta \in [0, \pi)$, such that $\phi \in [-\pi/2, \pi/2]$.

We are interested in the total $R_\star\left(\phi\right)$ of all stars in a field. We sum each star's $R_\star\left(\phi\right)$ and normalize by correcting for the total RHT intensity and the total area sampled, as follows:

\beq
\hat{R}\left(\phi\right) = \frac{\sum\limits_{n_{\star}} R_\star(\phi)}{\frac{1}{n_{\theta}} \int\limits_{\theta} \int\limits_x \int\limits_y R\left(\theta, x, y\right) \, d\theta dx dy}\left(\frac{\Omega}{n_{\star}\pi r^2}\right)
\eeq

Where $n_{\star}$ is the number of stars sampled in the field, $r$ is the sampling radius around each star, and $\Omega$ is the total area in the field.

\subsection{RHT Angle Expectation Value}\label{expectval}
In this section we describe a point estimator that quantifies the direction of a given region of ${R(\theta)}$. We choose the region $R_\star\left(\theta\right)$. 

We compute the angle
\beq
\label{carrotrht}
\left<\theta_{RHT} \right>^{\prime}= \frac{1}{2} \arctan \left[ \frac{\int \sin(2{\theta}) R_\star(\theta) \, d\theta}{\int \cos(2{\theta}) R_\star(\theta)\,d\theta} \right]
\eeq
and find the equivalent value on the interval ${\theta \in [0, \pi)}$,
\beq
\left< \theta_{RHT} \right> = \pi - \mathrm{mod}(\left<\theta_{RHT} \right>^{\prime} + \pi, \pi).
\eeq
This is the RHT angle expectation value, a measure of the orientation of the gas around a particular star. To compare this to the starlight polarization angle $\theta^\star$, one can simply take the difference in the two values:

\beq
\label{phi_RHT}
\left<\phi_{RHT}\right> =  \left<\theta_{RHT} \right> - \theta^\star,
\eeq
where, again, this subtraction must take place on the domain $\theta \in [0, \pi)$, such that $\left<\phi_{RHT}\right> \in [-\pi/2, \pi/2]$. Thus, if $\left<\phi_{RHT}\right> \simeq 0$, ${R_\star(\theta)}$ is well aligned with its starlight polarization angle.  \\

\subsection{RHT Distribution Widths}\label{distwidth}
The RHT angle expectation value ${\left<\theta_{RHT} \right>}$ is a useful metric for generalizing the orientation of the gas, but ignores all information about the strength and shape of $R_\star\left(\theta\right)$. A narrowly peaked $R_\star\left(\theta\right)$ and a much broader $R_\star\left(\theta\right)$ can have the same $\left<\theta_{RHT} \right>$. Similarly, the amount of intensity detected by the RHT is ignored in calculating $\left<\theta_{RHT} \right>$, which could be useful in determining the certainty of our angle estimation.

Another approach is to characterize the spread in the distribution. We report the interquartile range IQR$({\hat{R}(\phi)})$ as a metric for the width of ${\hat{R}(\phi)}$. One can also report the IQR of the $\left<\phi_{RHT}\right>$ measures around all stars in a field, IQR${\left(\left<\phi_{RHT}\right>\right)}$.

We note that IQR$({\hat{R}(\phi)})$ and IQR${\left(\left<\phi_{RHT}\right>\right)}$ for a collection of stars are qualitatively different metrics. Assuming the linear structure is aligned with the magnetic field as traced by starlight polarization in a given region, the IQR${\left(\left<\phi_{RHT} \right>\right)}$ reports how accurately one could predict the starlight polarization angle for the given RHT data, independent of RHT intensity. The IQR$({\hat{R}(\phi)})$ is inherently biased toward higher RHT intensity, and thus is a measure of how well one could predict the starlight polarization angle weighted by RHT intensity. Thus, if the strength of the RHT is a measure of the reliability of our prediction of the starlight polarization angles, IQR$({\hat{R}(\phi)})$ will typically be narrower than IQR${\left(\left<\phi_{RHT}\right>\right)}$.

\section{Data}\label{data}
We present an analysis of diffuse \hi from two surveys, each sensitive to a broad range of spatial scales. The Galactic Arecibo L-Band Feed Array \hi Survey \citep[GALFA-\hia;][]{2011ApJS..194...20P} maps 13,000 deg${^2}$ with 4${\arcmin}$ spatial resolution, 0.18 km s${^{-1}}$ spectral resolution, and ${\sim}$60 mK rms brightness temperature noise for a 1 km s$^{-1}$ velocity bin. We analyze a region of sky with 115.0${\degree}$ ${\le}$ RA ${\le}$ 245.0${\degree}$, and 23.0${\degree}$ ${\le}$ ${\delta}$ ${\le}$ 33.0${\degree}$: a 1,300 deg${^2}$ region of sky relatively devoid of telescope scan artifacts in the first data release, DR1. This is a strip of sky from $l$, $b$ $\sim$ (45$\degree$, 45$\degree$) to (190$\degree$, 20$\degree$) that encompasses Galactic zenith. 
For GALFA-\hi data we present an analysis of the velocity range from -7.0 km s$^{-1}$ to -1.1 km s$^{-1}$ where the fibers are most evident. We note that modifying this velocity range does not dramatically change the RHT-starlight polarization correlation. 

The second survey we analyze is the Parkes Galactic All Sky Survey (GASS; \citei{McClureGriffiths:2009dn}).  GASS maps the southern celestial sky at all declinations ${\delta}$ ${\le}$ 1${\degree}$ with 16${\arcmin}$ spatial resolution, 1 km s${^{-1}}$ spectral resolution, and 57 mK rms brightness temperature noise per 1 km s${^{-1}}$ channel.  We analyze the entire spatial area of GASS, excluding the region ${\left| b \right| <}$ 30${\degree}$ to remain focused on high latitude features.  GASS data is analyzed from 1.6 km s$^{-1}$ to 5.8 km s$^{-1}$. Again, the RHT-starlight polarization correlation is insensitive to the exact velocity range. 

In addition to these two surveys, we present an analysis of an \hi cold cloud in the Galactic plane with previously identified filaments in \S \ref{rccloud}. Observations of the Riegel-Crutcher cloud were obtained by \cite[][hereafter McC-G06]{McClureGriffiths:2006wx} as part of an extension to the Southern Galactic Plane Survey (\citei{McClureGriffiths:2005dv}). The data have a resolution of 100${\arcsec}$ (0.06 pc at the 125 pc distance of the cloud) and a channel spacing of 0.82 km s$^{-1}$. The data are analyzed from $+3.30$ km s$^{-1}$ to $+7.42$ km s$^{-1}$. 

The starlight polarization data corresponding to the GASS survey area are from the Heiles (\citeyear{Heiles:2000un}) compilation, an aggregation of starlight polarization catalogs that contains 9,286 stars. In the GALFA-\hi region, the \citeauthor{Heiles:2000un} (\citeyear{Heiles:2000un}) compilation is supplemented with stars from \cite{Berdyugin:2001cv} and \cite{2002A&A...384.1050B} which catalog 336 stars and 116 stars in the region of the North Galactic Pole, respectively. In cases where the same star is measured in more than one catalog, we defer to the more modern measurement. All catalogs contain optical measurements of starlight polarization angles. We exclude any stars in the catalogs that were part of targeted polarization studies of clusters, in order to have a star sample that is well distributed across the sky. We did not apply a distance or polarization intensity cut for the stars used with GALFA-\hi and GASS, though we did exclude stars with quoted errors on the starlight polarization angle greater than 25\degree. This leaves us with 153 stars in the GALFA-\hi region, and 3,206 stars in the GASS region. The stars used in the GALFA-\hi region have a median distance of 253 pc and an interquartile range of 133--442 pc.  The stars used for the GASS correlation have a median distance of 1140 pc and an interquartile range of 291-2218 pc.\\

\section{Fibers in Diffuse HI}\label{diffuse}

\begin{figure}[h!]
\centering
\includegraphics[scale=0.15]{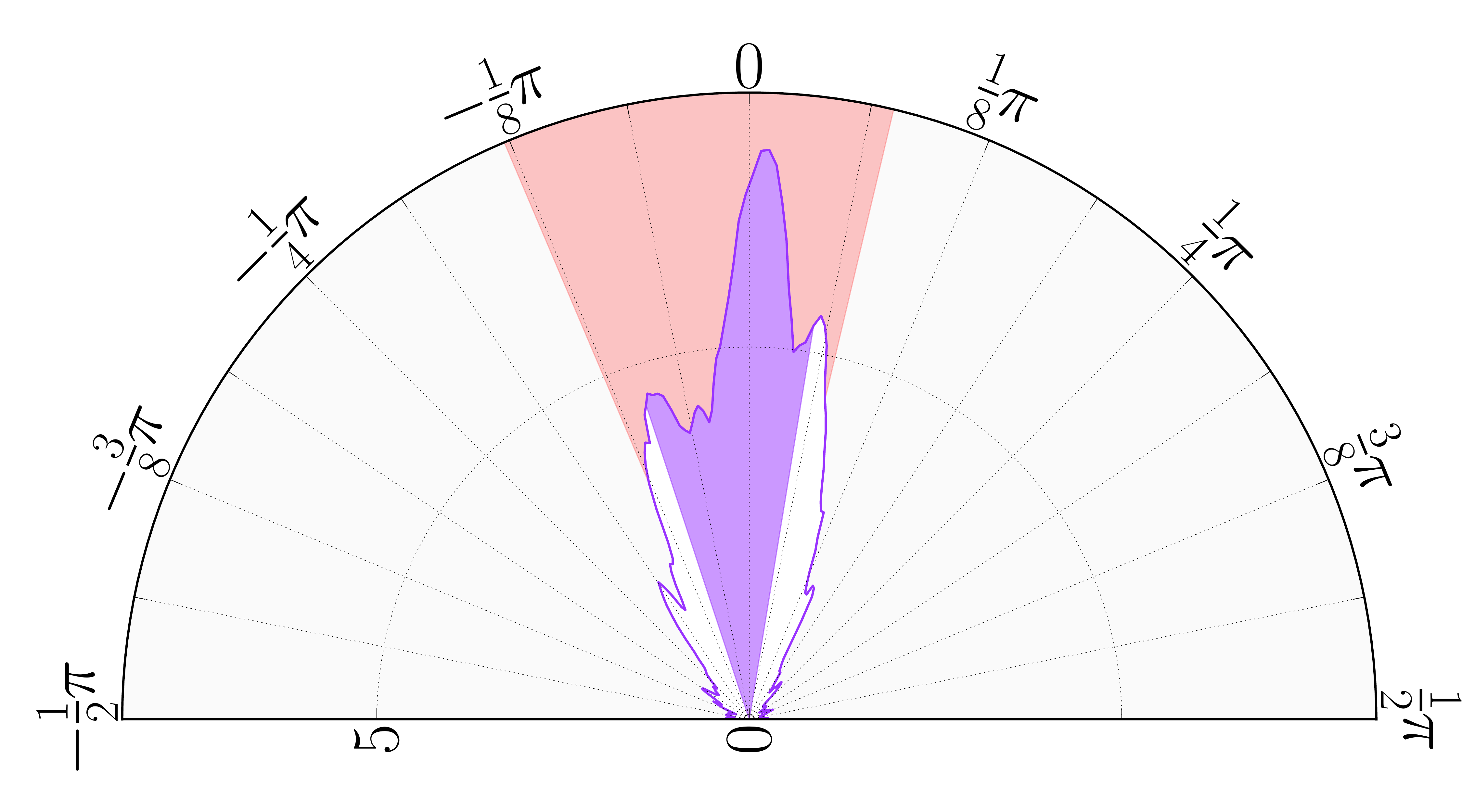}
\caption{Integrated RHT output ${\hat{R}\left(\phi\right)}$ (see \S\ref{starpolmethods}) for all stars in the GALFA-\hi field (purple line). The velocity range is -7.0 km s${^{-1}}$ to -1.1 km s${^{-1}}$, analyzed in two equal channels (see \S\ref{fiberparamspace}). The RHT was run with ($D_W, D_K, Z$) = (100$\arcmin$, 10$\arcmin$, 70\%). $R(\theta, x, y)$ is sampled in regions of radius 0.5$\degree$ around each star.  IQR$(\hat{R}\left(\phi\right))$ is 27\degree~ (purple shading). IQR${(\left<\phi_{RHT}\right>)}$ is 37\degree~ (red shading).\label{galfamain}}
\end{figure}

\begin{figure}[h!]
\centering
\includegraphics[scale=0.15]{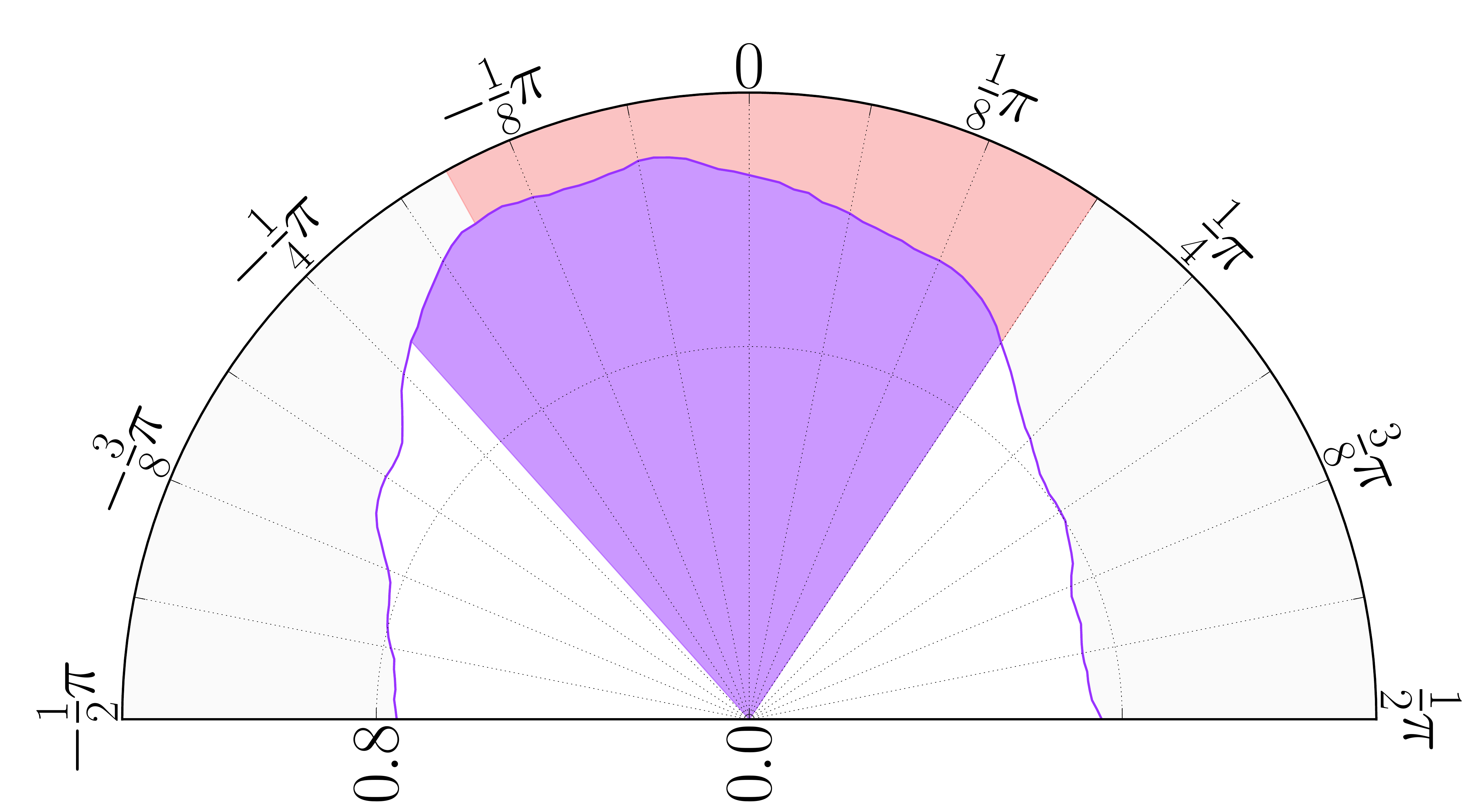}
\caption{Same as Figure \ref{galfamain}, but ${\hat{R}\left(\phi\right)}$ for all stars with ${\left| b \right| >}$ 30${\degree}$ in the GASS field. The RHT was run on data integrated over the velocity range 1.6 km s${^{-1}}$ to 5.8 km s${^{-1}}$. The RHT was run with ($D_W, D_K, Z$) = (245$\arcmin$, 53$\arcmin$, 70\%). $R(\theta, x, y)$ is sampled in regions of radius 2$\degree$ around each star. IQR${(\hat{R}\left(\phi\right))}$ is 77\degree~ (purple shading). IQR${(\left<\phi_{RHT}\right>)}$ is 65\degree~ (red shading).\label{gassmain}}
\end{figure}

\begin{figure}[h!]
\centering
\includegraphics[scale=0.35]{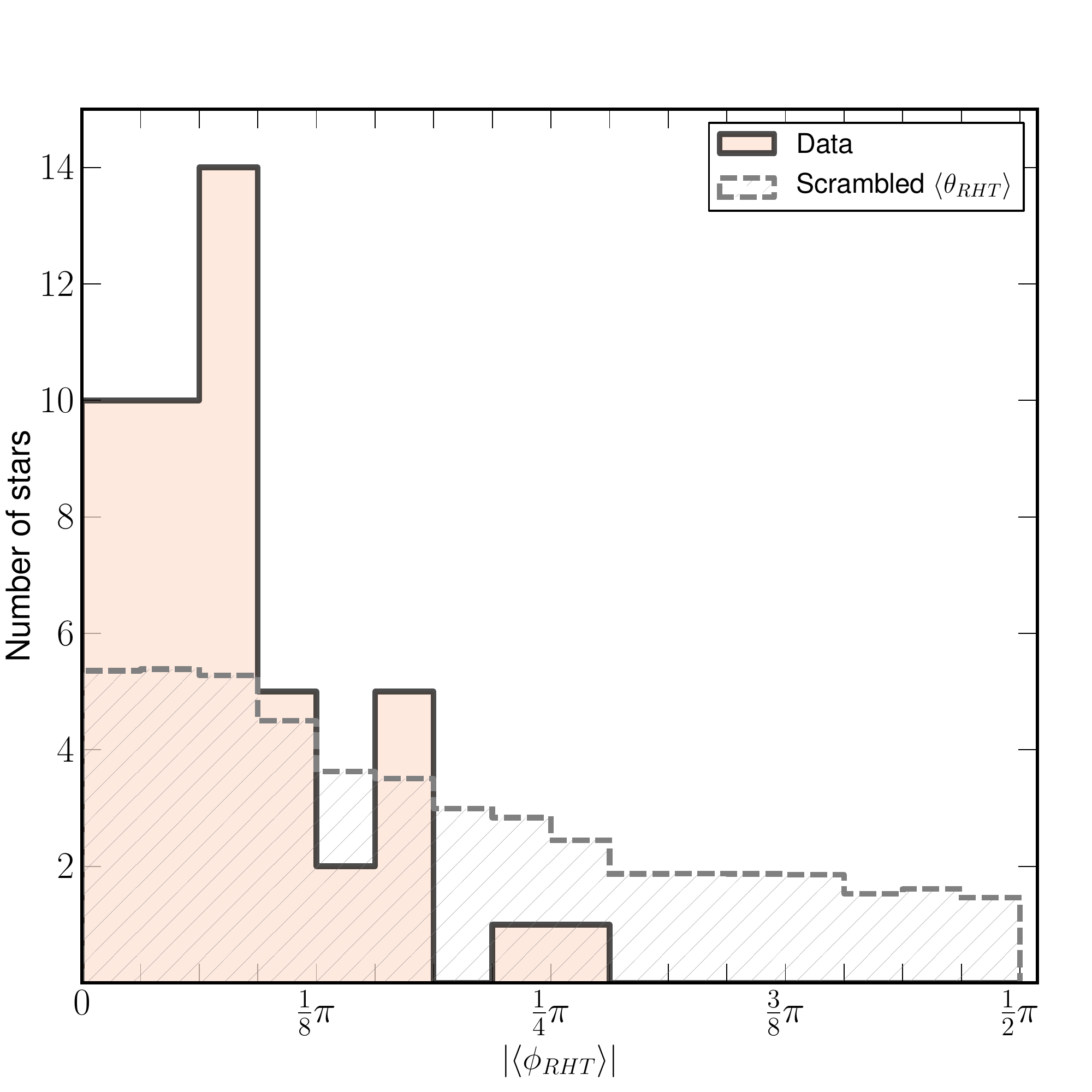}
\caption{Histograms of the difference between the measured starlight polarization angle and the RHT angle expectation value for all GALFA-\hi stars in the top quintile of RHT intensity (filled pink) and the same stars with scrambled $\left< \theta_{RHT} \right>$ values (hatched grey). Scrambled histogram is an average of $10^4$ random samples of $\left< \theta_{RHT} \right>$. The RHT-starlight polarization correlation is highly statistically significant (p $< 0.0001$). See \S\ref{corrstarpol}.}
\label{angdif_hists}
\end{figure}

\begin{figure*}[t!]
\centering
\includegraphics[width=0.9\textwidth]{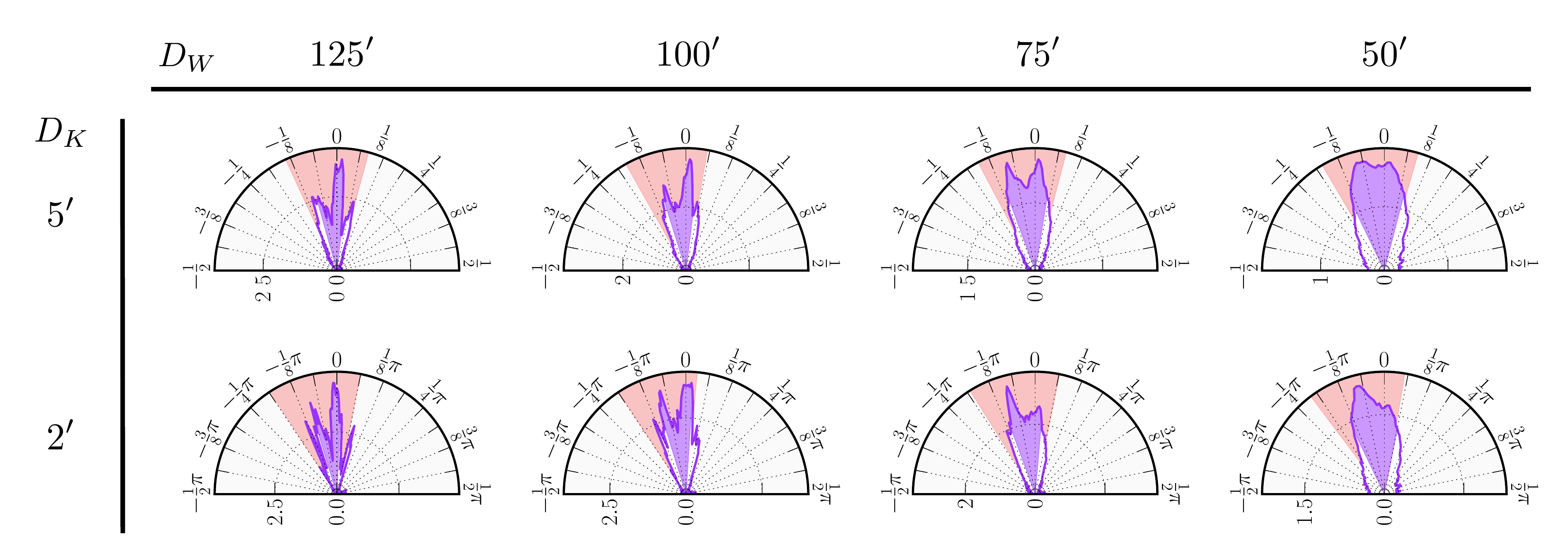}
\caption{A sample of the parameter space for GALFA. Smoothing kernel diameter ($D_K$) and window diameter ($D_W$) are indicated. All runs use an intensity threshold $Z$ = 70\%. Red shading indicates IQR$\left(\left<\phi_{RHT} \right>\right)$, purple shading indicates IQR${(\hat{R}\left(\phi\right))}$.}
\label{pspace_fig}
\end{figure*}

\begin{figure*}[h]
\centering
\includegraphics[width=\textwidth, scale=0.6]{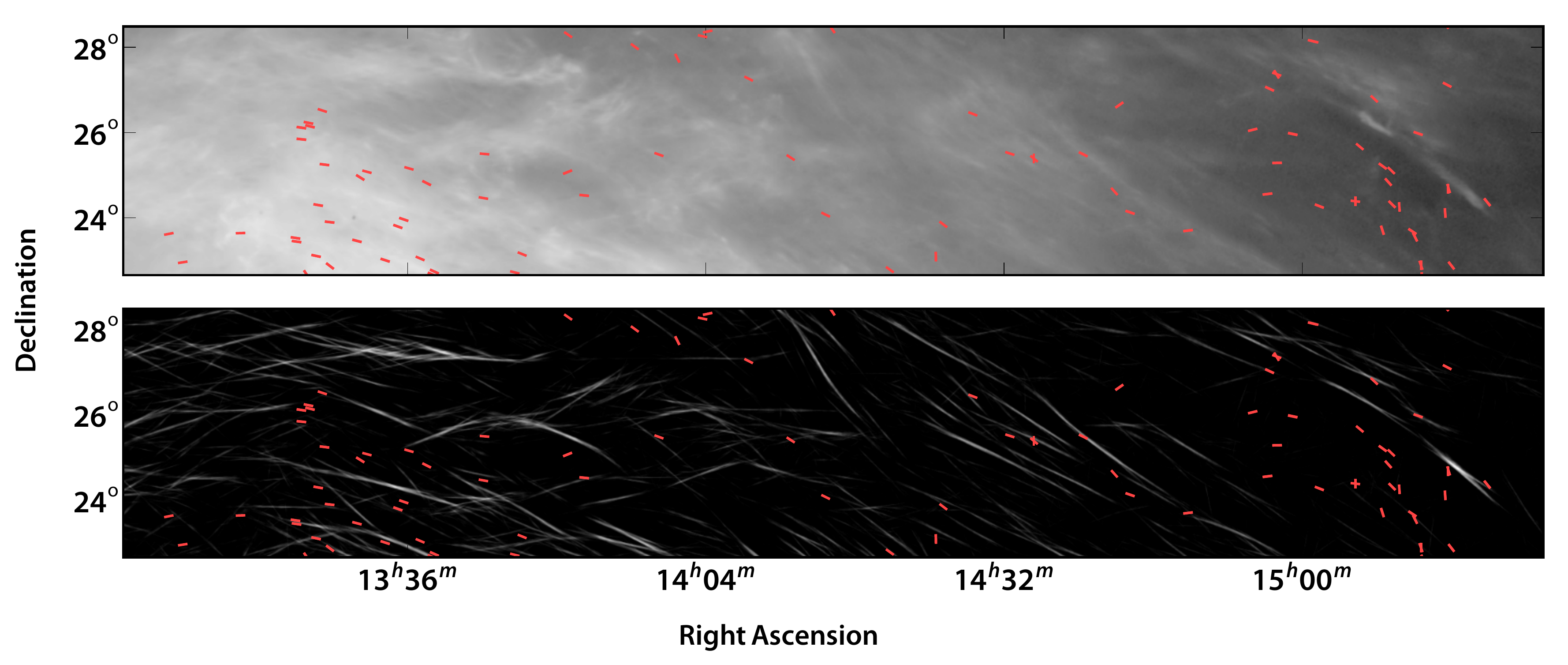}
\caption{A representative region of the GALFA-\hi data analyzed in \S\ref{diffuse}, shown in \hi emission (top) and RHT backprojection $R\left(x, y\right)$ (bottom; see Equation \ref{Bxy}). The images are integrated over the velocity range -7.0 km s${^{-1}}$ to -1.1 km s${^{-1}}$. Overlaid pseudovectors represent polarization angle measurements from the \cite{Heiles:2000un}, \cite{Berdyugin:2001cv}, and \citep{2002A&A...384.1050B} catalogs. In the top panel, the intensity scale is linear in log$(N_{HI})$, where black represents a column density of $2 \times 10^{18}$ cm$^{-2}$, and white is $2 \times 10^{20}$ cm$^{-2}$.}
\label{galfareg}
\end{figure*}

\begin{figure*}[h]
\centering
\includegraphics[width=\textwidth, scale=0.5]{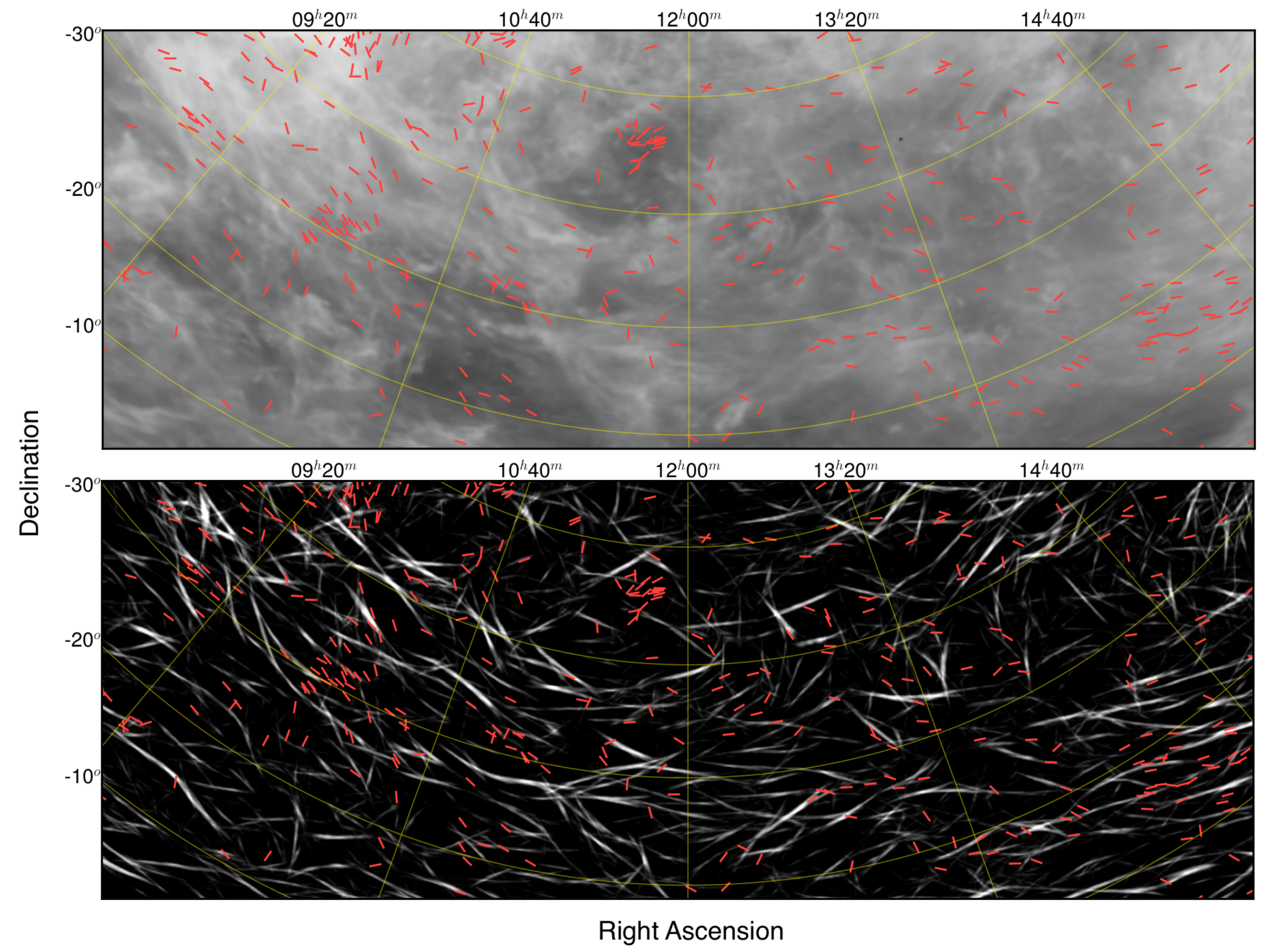}
\caption{A representative region of the GASS data analyzed in \S\ref{diffuse}, as in Figure \ref{galfareg}. The images are integrated over the velocity range 1.6 km s$^{-1}$ to 5.8 km s$^{-1}$. In the top panel, the intensity scale is linear in log$(N_{HI})$, where black represents a column density of $2 \times 10^{18}$ cm$^{-2}$, and white is $2 \times 10^{21}$ cm$^{-2}$.}
\label{gassreg}
\end{figure*}

\begin{figure*}[h!]
\centering
\includegraphics[width=\textwidth]{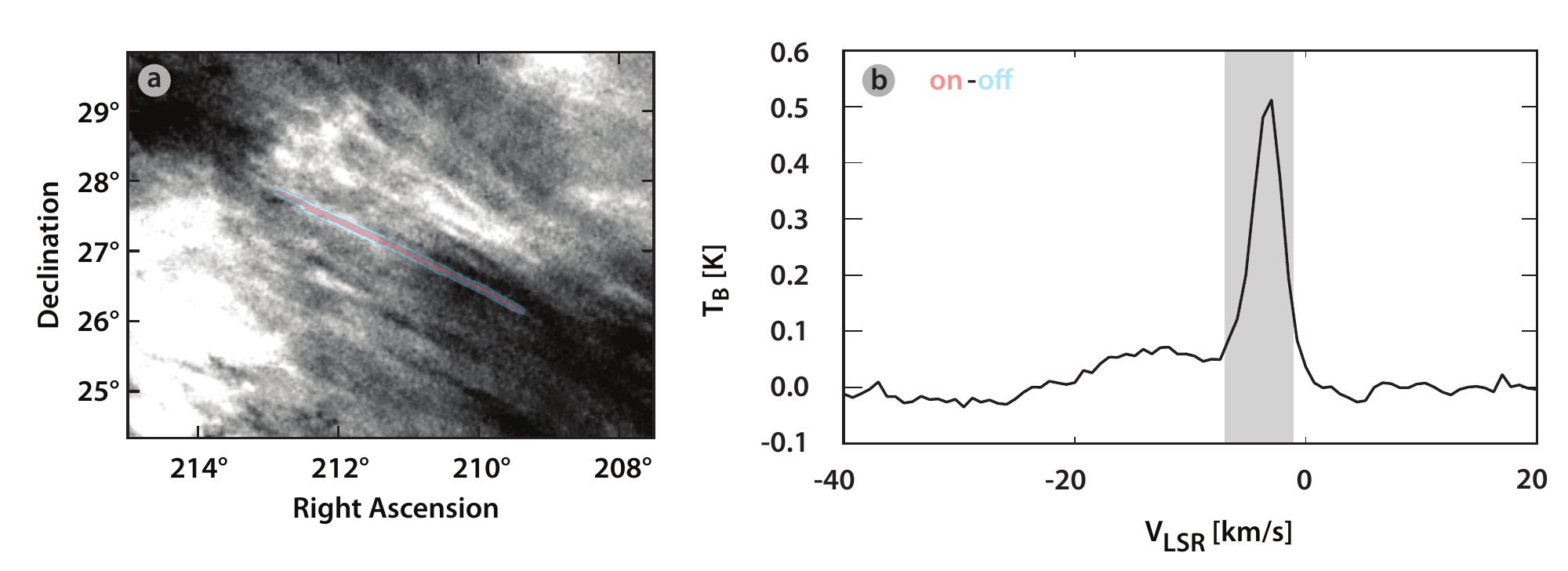}
\caption{(a) ``On'' fiber and ``off'' fiber fields overlaid on GALFA-\hi data. The image is integrated over the velocities indicated in (b). The fiber was selected from the RHT backprojection. Black represents a column density of $10^{19}$ cm$^{-2}$, white is $3 \times 10^{19}$ cm$^{-2}$. (b) The difference between the average spectrum in the on and off fields. Grey region indicates the velocity range analyzed for GALFA-\hi data, -7.0 km s$^{-1}$ to -1.1 km s$^{-1}$. See \S\ref{diffuse} for a discussion of fiber properties.}
\label{onoff}
\end{figure*}

\begin{figure*}[t!]
\centering
\includegraphics[width=0.9\textwidth]{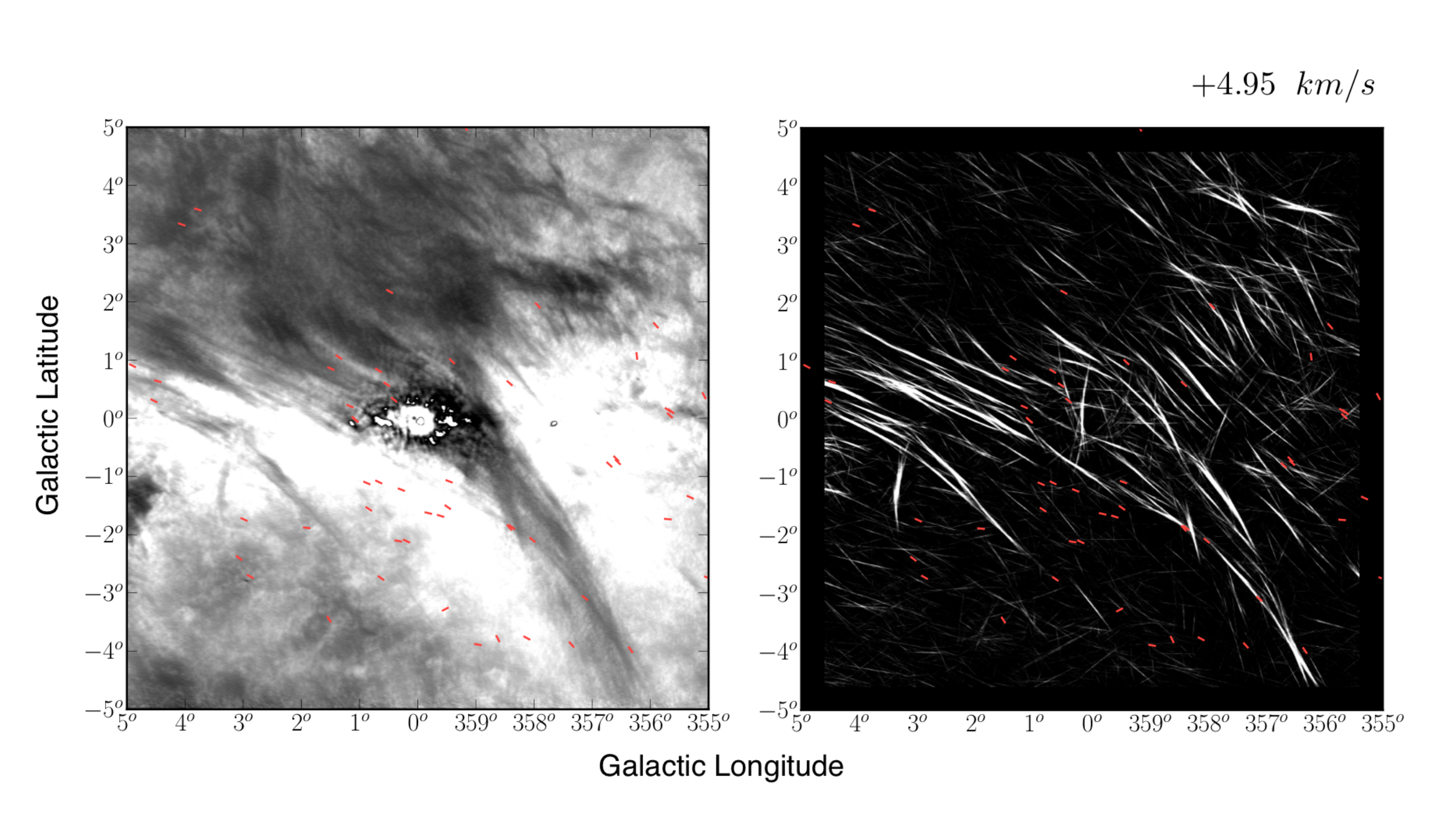}
\caption{The Riegel-Crutcher cloud (\S\ref{rccloud}) in \hi absorption (left) and RHT backprojection (right). Overlaid pseudovectors represent polarization angle measurements from the \citeauthor{Heiles:2000un} (\citeyear{Heiles:2000un}) compilation. In the left panel, the intensity scale is linear from -20 K (white) to -120 K (black).}
\label{rccloudfig}
\end{figure*}

\begin{figure}[bp]
\includegraphics[scale=0.24]{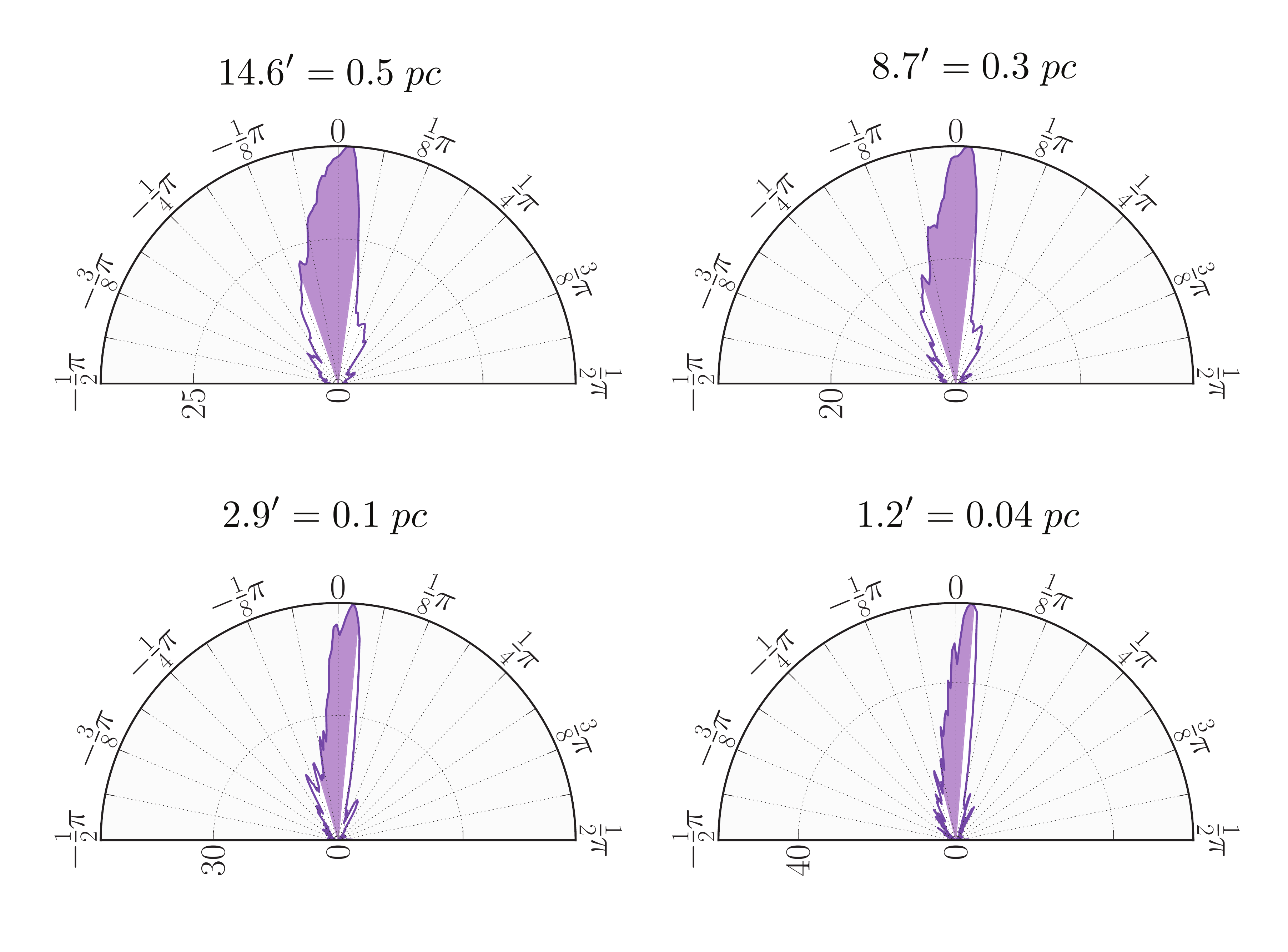}
\caption{${\hat{R}\left(\phi\right)}$ for all stars in the Riegel-Crutcher cloud (\S\ref{rccloud}). The radius of the sampling beam around each star is labeled above each figure, with sampling beam decreasing left to right from 14.6${\arcmin}$ to 1.2${\arcmin}$. Spatial radii of the sampling beams are calculated using the cloud distance of 125 pc. The width of the distribution decreases with decreasing beam size. As beam size decreases (top left to lower right): IQR$(\hat{R}\left(\phi\right))$ = 27.3$\degree$, 26.2$\degree$, 22.9$\degree$, 19.9$\degree$.}
\label{rcbeams}
\end{figure}

\begin{figure*}
\centering
\includegraphics[width=\textwidth, scale=0.5]{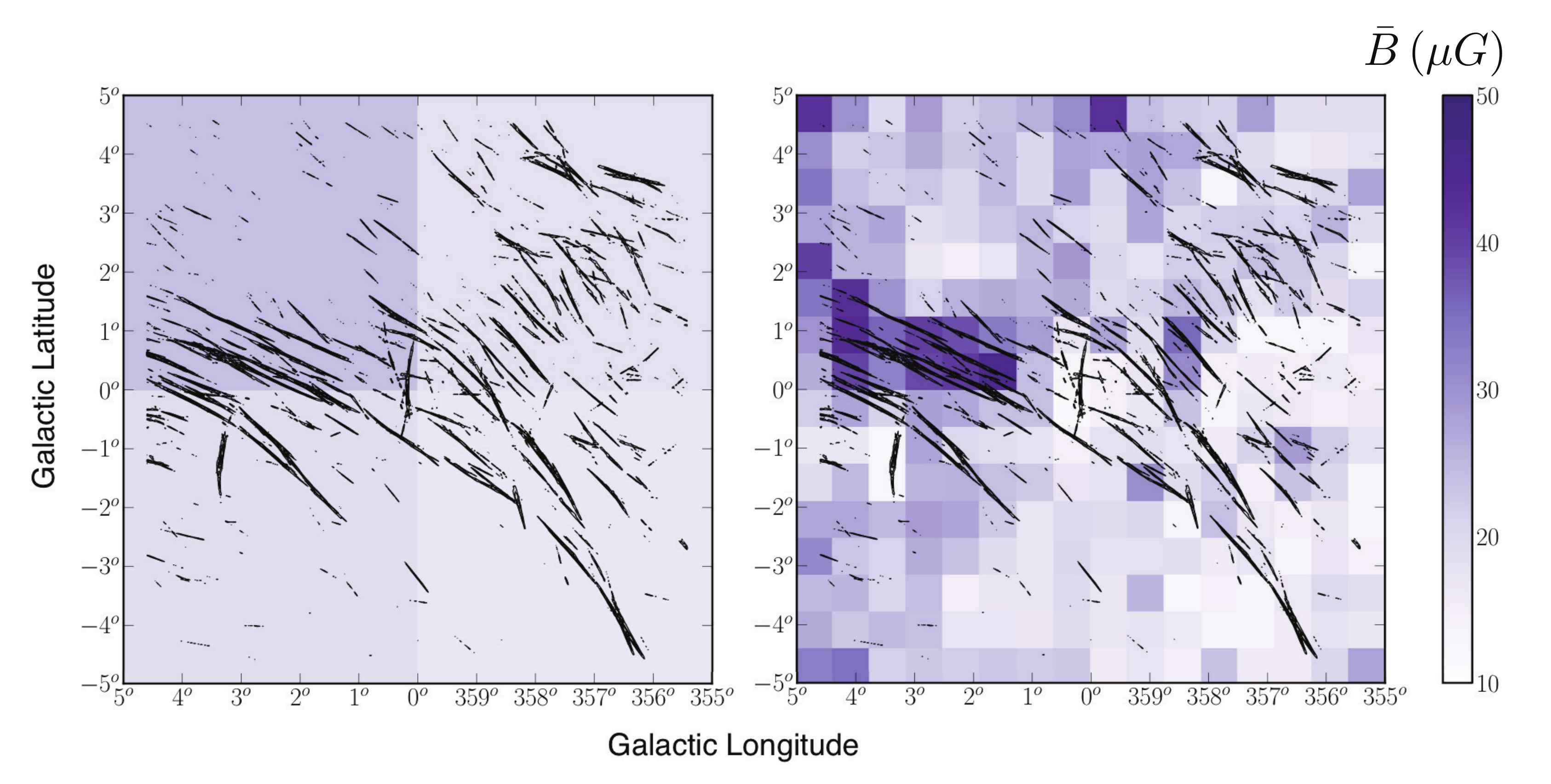}
\caption{Mean magnetic field strength ${B_{RHT}}$ calculated using the modified Chandrasekhar-Fermi method (\S\ref{cfmethod}) for 4 and 256 sections of the Riegel-Crutcher cloud. Density contours of the RHT backprojection are overlaid to give an idea of the fiber geometry (see Figure \ref{rccloudfig}).}
\label{cfcloudfig}
\end{figure*}

\begin{figure*}[h!]
\centering
\includegraphics[width=\textwidth, scale=0.5]{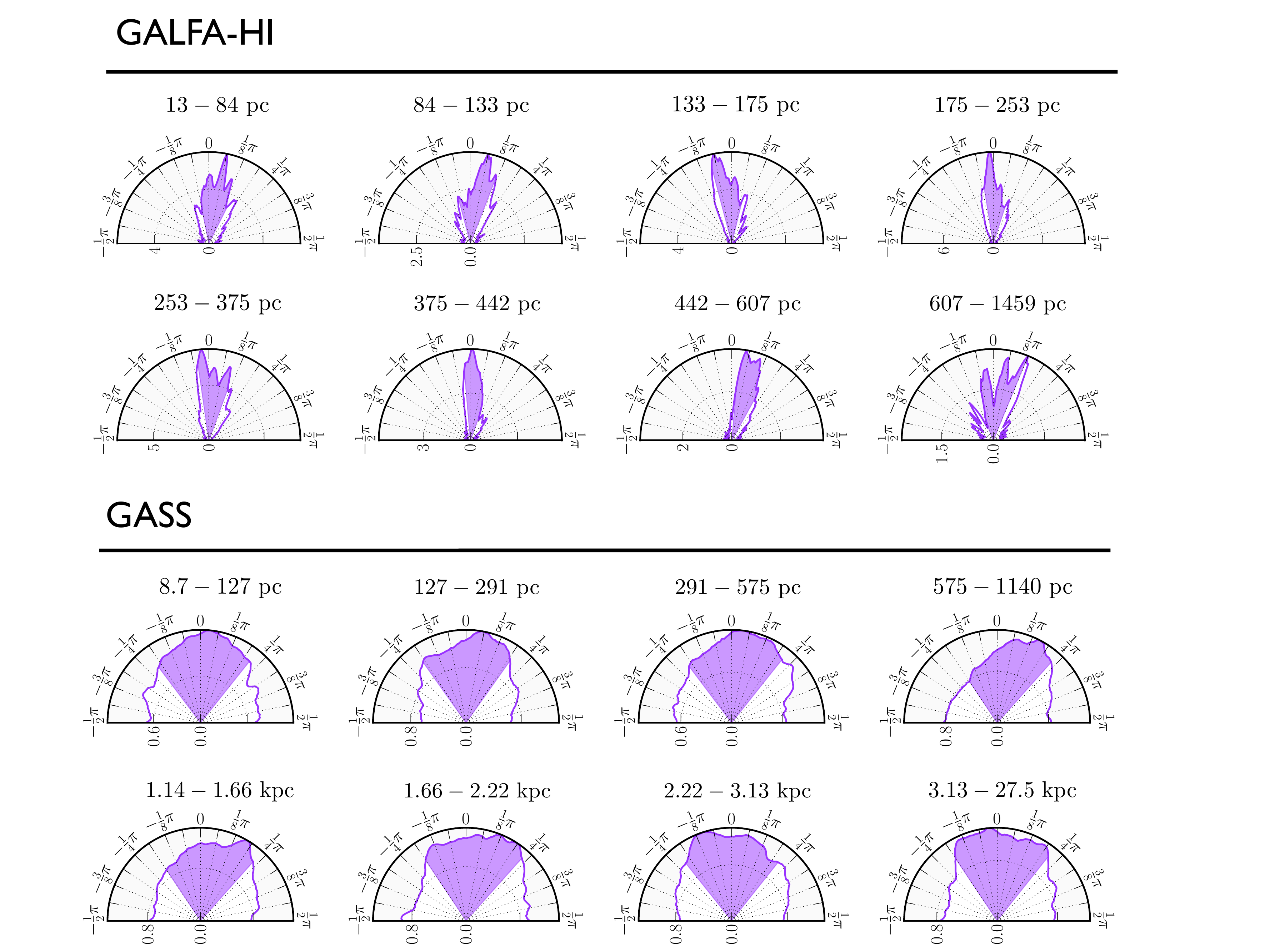}
\caption{${\hat{R}\left(\phi\right)}$ binned in star distance octiles for GALFA-\hi and GASS data (see \S \ref{diffuse}). There are approximately 18 stars in each GALFA-\hi distance octile, and 394 stars in each GASS distance octile. The median values of the $\hat{R}\left(\phi\right)$ distributions are consistent with a random sampling of distances.}
\label{octoplot}
\end{figure*}

Fibers in the diffuse, high latitude \hi are examined using the GALFA-\hi and GASS data sets. We find that $\hat{R}\left(\phi\right)$ is well-centered on zero in both data sets, with IQR$(\hat{R}\left(\phi\right)) = 27\degree$ for GALFA-\hi and 77$\degree$ for GASS (Figures \ref{galfamain} and \ref{gassmain}). 
These IQR$(\hat{R}\left(\phi\right))$ are measured for $(D_W, D_K, Z) = (100\arcmin, 10\arcmin, 70\%)$ and sampling radius $r = 0.5\degree$ for GALFA-H{\sc i}, and ($D_W, D_K, Z$) = (245$\arcmin$, 53$\arcmin$, 70\%) and sampling radius $r = 2\degree$ for GASS. This correlation indicates that the magnetic field is indeed aligned with the observed fibers. This result is robust to variation in RHT parameters (see Figure \ref{pspace_fig}). 

The alignment between $R_\star\left(\theta\right)$ and $\theta^\star$ in both GASS and GALFA-\hi data suggests that \hia-magnetic field alignment is a pervasive feature of the high-latitude ISM. However, fibers are not a scale-independent feature of the ISM; detection of the \hia-magnetic field alignment is much improved with better spatial resolution. By eye, and in the backprojection, the slender fibers in GALFA-\hi (Figure \ref{galfareg}) are largely absent from GASS (Figure \ref{gassreg}). This point is echoed by our study of the Riegel-Crutcher cloud in \S\ref{rccloud}.

\subsection{Parameter Space}\label{fiberparamspace}

We conduct a thorough exploration of the parameter space for the GALFA-\hi data. Rolling window diameters ($D_W$) from 50$^\prime$ to 125$^\prime$, smoothing kernel diameters ($D_K$) from 2$^\prime$ to 10$^\prime$, and intensity thresholds from Z = 50\% to 90\% (see Figure \ref{diagram}) were applied to the GALFA-\hi data. All combinations of parameters visually identify the same linear features in backprojection, and every $\hat{R}\left(\phi\right)$ displays a strong correlation with starlight polarization. This correlation is therefore robust to the variation of the RHT input parameters. Low intensity thresholds are computationally expensive because they require the storage of uniform background intensity. We select Z = 70\% for the duration of this work because lower intensity thresholds find the same linear features but store too much low-intensity background. Variation of the sampling radius $r$ does not significantly alter the observed RHT-starlight polarization correlation.

Figure \ref{pspace_fig} shows a representative sampling of the parameter space. We find that increasing $D_W$ narrows IQR${({\hat{R}\left(\phi\right)})}$, indicating that the longest, most linear features are the most well aligned with starlight polarization. However, IQR${\left(\left<\phi_{RHT} \right>\right)}$ remains consistent across parameter space, as this metric gives equal weight to the RHT-starlight polarization alignment around each star, regardless of RHT intensity. 

We also explore the effects of data channelization, $\delta v$. In the GASS data, we find the alignment is insensitive to whether we bin the data in advance of the RHT, or sum the $R_\star\left(\theta\right)$ from each channel. In the case of GALFA-\hia, it is possible to bin the data so finely (0.18 km s${^{-1}}$) that noise washes out the observed fibers, or to integrate over so many channels that fibers are less visually evident. In these cases the signal is detectably diminished. We split the velocity range -7.0 km s${^{-1}}$ to -1.1 km s${^{-1}}$ into two channels, run the RHT on each, and sum the $R_\star\left(\theta\right)$ from each channel, though the result is not sensitive to the exact channelization. 

\subsection{Correlation with Starlight Polarization}\label{corrstarpol}

We examine the star-by-star correlation between the measured starlight polarization angles $\theta^\star$ and the RHT angle expectation value $\left<\theta_{RHT}\right>$ in GALFA-\hi (Figure \ref{angdif_hists}). This allows us to determine whether the RHT-starlight polarization correlation exists on a fine scale, or simply in the large-scale orientation of the stars and gas, and to test the correlation robustness. We select the stars that sample regions in the top quintile of RHT intensity in each velocity channel ($\delta v$ from above), 48 stars in total. We expect the regions with the strongest RHT intensity to trace the most visually evident fibers. We calculate $\left|\left<\phi_{RHT}\right>\right|$ for each star (Equation \ref{phi_RHT}). We then scramble the $\left<\theta_{RHT}\right>$ values and recompute $\left|\left<\phi_{RHT}\right>\right|$ for each star. The scrambling is performed $10^4$ times and the results are averaged. The scrambled angle differences exhibit only a slight skew toward zero, indicating only a slight large-scale trend in fiber orientation. The unscrambled data is sharply skewed toward zero. The RHT-starlight polarization correlation is determined by a Monte Carlo analysis of the median to be highly statistically significant (p $< 0.0001$).

\subsection{Fiber Properties}\label{fiberprop}

We measure the properties of a GALFA-\hi fiber highlighted by the RHT backprojection. We note that the exact boundary of the fiber is dependent on the RHT input parameters, and that measured properties depend on the interpretation of the fiber as a distinct physical structure. 
We use the RHT backprojection to mask an ``on'' fiber and ``off'' fiber region, each of equal area, on the sky. Figure \ref{onoff} shows the average on minus average off spectrum and the selected regions of sky. We determine the line width of the spectrum to be 3.4 km s$^{-1}$ (FWHM) using a Gaussian fit. This fiber has a column density of $5.3 \times 10^{18}$ cm$^{-2}$, roughly typical of the GALFA-\hi fibers.

It is worth noting here that the column densities of the fibers discussed above are far too low to create the measured starlight polarization. To induce starlight polarization that can be measured accurately in the \citeauthor{Heiles:2000un} (\citeyear{Heiles:2000un}) catalog, a selective extinction of $\sim0.01$ is needed, equivalent to a column of $\sim$5$ \times 10^{19}$ cm$^{-2}$. Given the correlation between the magnetic field orientation and the fiber orientation, the fibers must be features of (or objects within) a dusty medium with a coaligned magnetic field, rather than the only elements of the medium. Indeed, because the starlight polarization angle represents the cumulative polarization of all material between observer and star, the discovery that the fiber orientation is correlated with the magnetic field orientation indicates that the fibers trace a structure that is co-aligned for a significant fraction of the dust along the line of sight. This correlation is discussed in the context of the local ISM morphology in \S \ref{lcwall}.

\section{Fibers in the Riegel-Crutcher Cloud}\label{rccloud}

The RHT can be applied to many different environments. We apply the method to a region of cold neutral medium: the Riegel-Crutcher cloud, an \hi self-absorption (HISA) feature at 125 pc toward the Galactic center (\citei{Heeschen:1955wi}; \citei{Riegel:1972wk}). The cloud was mapped in high resolution (100${\arcsec}$ = 0.06 pc at 125 pc) in McC-G06, who first resolved its exquisite filamentary structure and characterized the region as magnetically dominated. In the same work, the authors comment on the visibly apparent alignment of starlight polarization pseudovectors in the plane of the sky with the linearly elongated HISA structure. The RHT allows us to quantify this alignment.

Figure \ref{rccloudfig} shows the Riegel-Crutcher (hereafter R-C) cloud with polarization pseudovectors from the \cite{Heiles:2000un} catalog overlaid. Following McC-G06, we include all stars with ${-5 \degree < l < +5 \degree}$ and ${-5 \degree < b < +5 \degree}$, distances of less than 2 kpc, and polarization intensities of greater than ${1\%}$. This leaves 56 stars in the region. 

As the R-C cloud is composed of many thin linear features that are believed to be dominantly shaped by magnetic forces, the RHT-starlight polarization correlation should be very strong. Indeed, the degree of alignment is striking for a broad range of RHT input parameters. We run the RHT for a single velocity channel at a time to preserve all velocity information. All channels individually show strong RHT-starlight polarization alignment. Each panel in Figure \ref{rcbeams} shows ${\hat{R}\left(\phi\right)}$ for the velocity channels +3.30 km s${^{-1} \le v \le +7.42}$ km s${^{-1}}$, a range that encompasses the cloud visually (again following McC-G06).

The sharp alignment of ${R}_\star\left(\theta\right)$ with $\theta^\star$ in Figure \ref{rcbeams} demonstrates that the filaments trace the magnetic field, as expected. As the radius of the sampling beam decreases, IQR$(\hat{R}\left(\phi\right))$ decreases. For sampling beam radii of (14.6\arcmin, 8.7\arcmin, 2.9\arcmin, 1.2\arcmin), we find IQR$(\hat{R}\left(\phi\right))$ = (27.3$\degree$, 26.2$\degree$, 22.9$\degree$, 19.9$\degree$). As the alignment is significantly better with a smaller sampling beam for $R\left(\theta, x, y\right)$, we infer that the RHT is not simply confirming the evident large-scale orientation of the magnetic field, but actually tracing the fine magnetic structure in the region. We have checked and confirmed that the alignment of $\hat{R}\left(\phi\right)$ is not dominated by a few stars. \\

\section{Toward A Resolved Chandrasekhar-Fermi Method}\label{cfmethod}
The result that the RHT traces small-scale variation in the magnetic field in the R-C suggests that the RHT may provide a reasonable proxy for starlight polarization measurements in regions where the RHT and starlight polarization are in close alignment. For such regions we propose an extension of the Chandrasekhar-Fermi method for estimating the magnetic field strength in the plane of the sky.

Originally proposed by Chandrasekhar and Fermi (\citeyear{CHANDRASEKHAR:1953uq}) to estimate the field strength in spiral arms, the Chandrasekhar-Fermi method uses starlight polarization to estimate the average field strength ${\left< B \right>}$ in a region. The method relates the line-of-sight velocity dispersion (${v_{los}}$) to the dispersion of starlight polarization angles about a mean component. Assuming that turbulence isotropically randomizes the magnetic field in the region, the mean field strength is given by
\beq
B_{CF}^2 \equiv \overline{B}^2 = \xi 4\pi\rho \, \frac{\sigma(v_{los})^2}{\sigma(\tan (\delta^\star))^2},
\eeq
where
\beq
\delta^\star \equiv \theta^\star - \overline{\theta^\star },
\eeq
${\rho}$ is the gas density,  $\overline{\theta^\star }$ is the mean starlight polarization angle, and ${\xi}$ is a correction factor representing the ratio of turbulent magnetic to turbulent kinetic energy (e.g. \citei{Heitsch:2001da}). The validity of the method depends critically on the presence of a significant mean field component.

We apply a modified Chandrasekhar-Fermi method to the R-C cloud described in \S\ref{rccloud}. Following McC-G06, we adopt ${\xi = 0.5}$, ${\rho = 1.4m_Hn_H =}$ 1.1 x 10${^{-21}}$ g cm${^{-3}}$, and ${\sigma_{v_{los}} = \sigma_{turb}}$ = 1.4 km s${^{-1}}$. Instead of $\theta^\star$ we substitute the expectation value of the RHT evaluated at every pixel in the image, ${\left<\theta_{RHT}\right>}_{\rm pixel}$, where ${\left<\theta_{RHT}\right>}_{\rm pixel}$ is the equivalent of  ${\left<\theta_{RHT}\right>}$, substituting $R\left(\theta, x_0, y_0\right)$ for $R_\star\left(\theta\right)$ in Equation \ref{carrotrht}. Thus we are evaluating:
\beq
B_{RHT}^2 \equiv \overline{B}^2 = \xi 4\pi\rho \, \frac{\sigma(v_{los})^2}{\sigma(tan (\delta_{RHT}))^2}
\eeq
where
\beq\label{deltarht_pixel}
\delta_{RHT} \equiv \left<\theta_{RHT}\right>_{\rm pixel} - \overline{\left<\theta_{RHT}\right>_{\rm pixel}}
\eeq
Because we obtain a ${\left<\theta_{RHT}\right>_{\rm pixel}}$ value for every pixel in the image space, ${\overline{\left<\theta_{RHT}\right>_{\rm pixel}}}$ can be evaluated over a region of any size that contains significant RHT signal.

Evaluating ${\overline{\left<\theta_{RHT}\right>_{\rm pixel}}}$ over the full extent of the R-C cloud, we obtain ${B_{RHT}}$ = 19 ${\mu}$G. McC-G06 report ${B_{CF}}$ = 60 ${\mu}$G for the region -3${\degree <}$ l ${\le}$ 5${\degree}$,  -3${\degree <}$ b ${\le}$ 5${\degree}$. In this same region, we obtain ${B_{RHT}}$ = 23 ${\mu}$G. Figure \ref{cfcloudfig} shows ${B_{RHT}}$ evaluated over smaller regions of sky, to demonstrate the possibility of a resolved Chandrasekhar-Fermi method. Each colored square in Figure \ref{cfcloudfig} represents $B_{RHT}$ calculated using all ${\left<\theta_{RHT}\right>_{\rm pixel}}$ values in that square. Pixels containing no RHT power are not included in the computation of Equation \ref{deltarht_pixel}. The strongest ${B_{RHT}}$ we find in a subregion of the cloud is ${\sim}$50 ${\mu}$G, near $(l, b) \sim (1.5\degree, 0.25\degree)$ (see Figure \ref{cfcloudfig}).

This should be considered a preliminary step in the development of a resolved Chandrasekhar-Fermi method. A thorough analysis of the limitations and error in the RHT point estimator $\left<\theta_{RHT}\right>_{\rm pixel}$, as well as testing with simulations, will be pursued in the future. Indeed we expect $\left<\theta_{RHT}\right>_{\rm pixel}$ to overestimate the true variability of magnetic field orientation, and thus underestimate $\bar{B}$. A weighting scheme based on $R_\star\left(\theta, x, y\right)$ would reduce this bias. We caution that the same assumptions hold as in the classical Chandrasekhar-Fermi method, in particular that a significant mean field element must be present for the field estimate to have meaning. Nevertheless, the na\"ive application of the method outlined here to the R-C cloud does achieve the same typical field-strength estimate as the classical Chandrasekhar-Fermi method.

\section{Discussion}\label{discussion}
The RHT is a powerful new tool for characterizing linear structure. This work quantifies for the first time the strong alignment between diffuse \hi fibers and the interstellar magnetic field. In this section we discuss the physical properties of the diffuse fibers, their relationship to the local cavity, and their significance in the context of modern magnetohydrodynamic simulations.

\subsection{Physical Properties of Fibers}
The GALFA-\hi and GASS surveys cover similar column density and latitude regimes, but differ by a factor of four in angular resolution (4${\arcmin}$ for GALFA, 16${\arcmin}$ for GASS). The strikingly collinear \hi fibers that prompted this investigation are visually evident in GALFA-\hi data (Figure \ref{galfareg}), and are not as apparent in the GASS data (Figure \ref{gassreg}). The fiber widths are in many cases visually unresolved even in the GALFA-\hi data, and so are on the order of or thinner than the GALFA-\hi spatial resolution. We find that the RHT-starlight polarization correlation is significantly higher in the GALFA-\hi data. Thus, the data are consistent with a model in which fine, magnetically aligned \hi fibers are ubiquitous in the high-latitude sky, but washed out at lower resolutions. 

The GALFA-\hi fibers have typical column densities that range from $\sim$10${^{19}}$ cm${^{-2}}$ down to our sensitivity limit of $\sim$10${^{18}}$ cm${^{-2}}$. A typical total Galactic \hi column density at high latitude is $\sim$3$\times 10{^{20}}$ cm${^{-2}}$, so an individual fiber does not dominate the column. Assuming a cylindrical geometry, we calculate an \hi volume density of n $\sim$ 14 cm$^{-3}$ for the fiber shown in Figure \ref{onoff}. 
If we interpret the linewidth as purely thermal, we find a temperature of 220 K; some of this linewidth may in fact be driven by turbulence within the cloud, so we consider this an upper limit on the temperature. The thermodynamic pressure is then $P/k_B$ = nT = $3200$ K cm$^{-3}$, consistent with the standard pressure found in the ISM at the solar circle \citep{Wolfire2003}. The angular length of the fiber identified by the RHT backprojection in Figure \ref{onoff} is about 5${\degree}$, although a typical fiber length is difficult to identify as they often exist in complexes of fibers up to 15${\degree}$ long and they may extend past the boundaries of the surveyed area. The physical scale of the fibers depends on the distance to the gas. If we choose a fiducial distance of 100 pc, the distance to the wall of the local cavity (\citei{Sfeir:1999tz}), the physical resolution of GALFA-\hi is 0.12 pc, and the length of the fiber in Figure \ref{onoff} is 8.7 pc.  As mentioned above, the widths are largely unresolved and therefore correspond to $<0.12$ pc for the GALFA-\hi fibers.  We investigate correlations between the polarization alignment and the location on the sky, extinction level, and polarization intensity and find no relationship. 

McC-G06 put a constraint on the magnetic field strength of the R-C cloud as $B_{tot} >$ 30 $\mu$G through the assumption that the magnetic energy density should dominate over the kinetic energy density to maintain the distinct linear nature of the filaments. If we apply this argument to a typical GALFA-\hi fiber we find ${B_{tot}} >$ 5 ${\mu}$G. This number is consistent with expectations for the magnetic field in the diffuse ISM \citep{Heiles:2005tb}. 

\subsection{Fibers and the Local Cavity Wall}\label{lcwall}

The Sun resides inside a largely evacuated volume of the ISM called the local cavity (LC). While the original theory that the LC is a bubble filled with hot, overpressurized, X-ray emitting gas has largely been overturned \citep{Koutroumpa09,WS10,2011ApJ...735..129P}, there is strong evidence that very little neutral gas and dust exists on this side of the LC wall, approximately 100 pc away (see \citei{2013arXiv1309.6100L} for a detailed map). In Figure \ref{octoplot} we show that the orientation of polarized starlight is well aligned with $\hat{R}\left(\phi\right)$, independent of the distance to the stars. Our stellar compilation only includes stars with relatively low errors in polarization angle measurement ($\Delta \theta < 25$\degree, see \S \ref{data}), which tend to have higher polarization percentages, and thus are behind more polarizing material. The median distance to stars $|b| > 30$\degree~ that meet this criterion is 144 pc, while the median distance to stars that fail this criterion is 43 pc. The wall of the LC is often defined as the distance at which $N_{HI} > 10^{19}$ cm$^{-2}$ \citep{CR87}, which is equivalent to an extinction $E\left(B-V\right)$ of only 0.002 \citep{2013ApJ...766L...6P}, too low to produce well-measured polarization angles in our compiled data set. Thus, essentially by definition, all of the stars we consider in this analysis are outside of the LC. If the fibers are a part of the wall of the LC, this explains why we do not see a marked decrease in correlation as we examine farther stars; they too are being polarized by the gas in the LC wall.

If there were a signficant column of dust-bearing gas beyond the LC wall, unaffected by the structure of the LC itself, it would presumably have a relatively uncorrelated magnetic field orientation. This would generate a decreased RHT-starlight polarization angle correlation. Since no such decorrelation is detected (Figure \ref{octoplot}), we find that the vast majority of the high Galactic latitude column is in or near the LC wall. This is consistent with modern tomographic maps of the local ISM \citep{Vergely10, 2013arXiv1309.6100L}. The thickness of the wall is not yet well constrained. We note that there may be a hint of decorrelation in the farthest distance bin in the GALFA-\hi data. This may be due to the presence of the intermediate-velocity arch, which covers much of the GALFA-\hi area and resides at approximately 1 kpc above the disk \citep{Kuntz:1996ff}.

This result points towards a formation and alignment mechanism for the fibers similar to that described in \citet{Weaver:1979vh} and further quantitatively developed in \citet{1998LNP...506..227H} for the Sco-Cen association and Radio Loop I. To paraphrase, many megayears ago a collection of massive stars produced outflows, and in the case of the LC, supernovae \citep{CR87}. These winds and explosions inflated a bubble of gas and dust and stretched the cavity wall to create the aligned fibers and magnetic field lines we detect. Whether or not this description fully explains the fibers and their magnetic alignment, any explanation must take into account the formation and structure of the local ISM.

\subsection{Simulations of Linear Structures}

Much of the simulation work linking gas morphology to magnetic field structure is focused on understanding molecular clouds. Observational evidence for magnetic influence on gravitational collapse includes regions where the magnetic field is oriented orthogonal to the densest structures in a molecular cloud, but parallel to the surrounding lower density medium, apparently owing to self-gravitational collapse along the field lines (e.g. \citei{Goldsmith:2008hl}, \citei{Nakamura:2008iv}). Recent \emph{Herschel} observations (\citei{Molinari:2010hm}, \citei{Peretto:2012bu}, \citei{Andre:2010ka}) have sparked an interest in modeling the formation of more diffuse molecular filaments, where a complex interplay between turbulence, gravity, and magnetism determine the alignment between filaments and magnetic fields. \citeauthor{Soler:2013dh} (\citeyear{Soler:2013dh}) modeled turbulent molecular clouds and found a link between the gas morphology and the orientation of the magnetic field. For diffuse, high-latitude \hia, gravity is unlikely to play a role in fiber formation and magnetic alignment.

Major progress has been made in this low density regime by \cite{Hennebelle:2013ws}, who showed that linear features can be created and maintained in a turbulent ISM without appealing to gravity. \cite{Arzoumanian:2011wu} detected a typical width for dust filaments of 0.1 pc in \emph{Herschel} data. In simulations conducted in \citeauthor{Hennebelle:2013ws} (\citeyear{Hennebelle:2013ws}; see also \citei{Hennebelle:2013va}) they reproduce this characteristic width in regions shielded from UV radiation, the scale being set by the dissipative process of ion-neutral friction.  Exposed, non-gravitating features, such as the fibers examined in this paper, are expected to have widths at least 10 times smaller due to higher ionization and lower densities. This prediction is consistent with our finding that we are increasingly resolving the fibers that are aligned with the magnetic field with higher resolution observations. The fibers are more apparent and better aligned with the field in the GALFA-\hi data than in the GASS data, or a resolution of 0.12 pc vs. 0.47 pc (at 100 pc), respectively.  For the \hi absorption filaments probed in the R-C cloud again the alignment improves as we decrease the radius of the sampling beam.  This is consistent with the width of the filaments largely being unresolved, or $<0.06$ pc at the cloud's distance of 125 pc. We can test the prediction of 0.01 pc wide fibers with yet higher resolution, highly sensitive observations enabled by instruments like the JVLA and the SKA pathfinder telescopes. 

\section{Conclusions}\label{conclusion}

This paper used \hi surveys of the Galactic ISM to study the relationship between gas morphology and the structure of the interstellar magnetic field. The highlights are summarized as follows.
\begin{itemize}
\item{We identified a novel set of features in the diffuse, high Galactic latitude \hi ISM: slender, linear, clustered features we call \hi fibers.}
\item{We developed a method for quantifying the coherent linearity of structures in images called the Rolling Hough Transform. }
\item{We used the RHT to demonstrate that the orientation of the fibers is correlated with the orientation of starlight polarization. This result is largely independent of the RHT input parameters $D_W, D_K,$ and $Z$, as well as velocity binning $\delta v$.}
\item{The magnetic fields and linear \hi features are aligned throughout the high Galactic latitude ISM, but this effect is not scale free.  Higher resolution observations show a much higher correlation between the fibers and the field.  The fibers are largely unresolved even with the highest resolution observations at 0.06 pc.}
\item{The GALFA-\hi and GASS fiber features are most likely a component of the local cavity wall and their derived physical properties at 100 pc are consistent with this environment.}
\item{We propose a technique based on the Chandrasekhar-Fermi method to measure the magnetic field strength in regions with strong, pervasive fields using only the RHT.}
\end{itemize}

The results of this work suggest a number of avenues for future exploration. The most obvious is to expand the work to larger areas of sky at higher resolution. In the northern celestial sky, EBHIS \citep{Kerp:2011ga} will provide a map similar to that of GASS with slightly higher resolution ($9^\prime$) and slightly lower sensitivity. The GALFA-\hi second data release will provide ten times more area at 4$^\prime$ resolution than the region examined here. In the future, SKA pathfinders APERTIF \citep{Verheijen:2009ua} and ASKAP \citep{Duffy:2012ie, 2012arXiv1207.0891D} will provide sub-arcminute resolution observations of the entire sky. The RHT can also be applied to observations of other phases of the magnetized ISM, for instance in molecular gas and dust, and likely would be an appropriate tool for any region not strongly dominated by gravity. Indeed, the RHT may even be a useful tool for finding stellar stream features in the Galactic halo. Furthermore, since the Hough transform can be generalized to find practically any template in the image plane \citep{Duda:1972uj}, the RHT could be extended to search for shells, cometary structures, or any other pervasive morphological feature of the ISM.

Another clear direction is the pursuit of comparable structures in simulations of the ISM. To date, we know of no examples in the ISM simulation literature in which magnetic fields are shown to be aligned with linear, neutral structures in diffuse media similar to that discussed here. This may be because multi-phase, magnetized simulations of a realistic Galactic ISM \citep[e.~g.~][]{2012arXiv1202.0552H} are never conducted at high enough resolution to resolve the features we detect. We suggest that a zoom-in of such a simulation near the Galactic disk at higher resolution or an implementation with an adaptive mesh (or both) may be able to resolve the \hi fibers. If simulations were to be unable to generate these kinds of features and correlations, it would suggest that the fibers are dependent on physics we are still incapable of capturing in simulations.

The discovery that the RHT can, at least in magnetically dominated regions, trace fine magnetic field structure, invites further investigation of the relationship between RHT angle dispersion and the magnetic field strength, and the efficacy of a resolved Chandrasekhar-Fermi method. To do this properly, we suggest the simulation work discussed above could be used to determine any bias or scaling that are needed to apply our method to other data accurately (as in \citei{Heitsch:2001da}).

\acknowledgements

The authors thank Destry Saul, Sne\v zana Stanimirovi\' c, Carl Heiles, Erik Rosolowsky, Patrick Hennebelle, Fabian Heitsch, and Fran\c cois Levrier for useful discussion, Naomi McClure-Griffiths for providing the Riegel-Crutcher cloud data, and the anonymous referee for many helpful comments. We also thank other members of the GALFA-\hi team and members of the ALFALFA team for their role in making the survey possible.
We acknowledge support from the Luce Foundation. S.E.C. was supported by a National Science Foundation Graduate Research Fellowship under Grant No. DGE-11-44155. J.E.G.P. was supported by HST-HF- 51295.01A, provided by NASA through a Hubble Fellowship grant from STScI, which is operated by AURA under NASA contract NAS5-26555.

\bibliographystyle{yahapj}

\end{document}